\def\la15{La$_{1.85}$Sr$_{0.15}$CuO$_4$}

\def\y7{YBa$_2$Cu$_3$O$_7$}
\def\y695{YBa$_2$Cu$_3$O$_{6.95}$}
\def\yde{YBa$_2$Cu$_3$O$_{7-\delta}$}

\def\dbc{DyBa$_2$Cu$_3$O$_{7-\delta}$}
\documentstyle[aps,prl,amssymb,amsfonts,epsf]{revtex}
\begin{document}
\draft

\twocolumn[\hsize\textwidth\columnwidth\hsize\csname @twocolumnfalse\endcsname

\title{Penetration Depth and Conductivity of NbN and \dbc~Thin Films
Measured by mm-Wave Transmission}

\author{B. J. Feenstra, F. C. Klaassen and D. van der Marel}
\address{Materials Science Centre,Laboratory of Solid State Physics,
University of Groningen, \mbox{Nijenborgh 4}, 
9747 AG Groningen, The Netherlands}

\author{Z. H. Barber}
\address{Dept. of Materials Science and Metallurgy, University of Cambridge,
Pembroke Street, Cambridge CB2 3QZ, United Kingdom}

\author{R. Perez Pinaya, M. Decroux}
\address{Dept. de Physique de la mati\a`ere condens\a'ee, University of Geneva, 
24, quai Ernest-Ansermet, CH-1211 Gen\a`eve 4, Switzerland}

\date{\today}
\maketitle
\begin{abstract}
Using mm-wave transmission we obtain information about both the 
real and imaginary part of the dielectric function of superconducting
thin films. This is done by fitting the Fabry-Perot resonance spectrum
for the film plus substrate over a broad frequency range (110-180 GHz)
using the full Fresnel equations and a two-fluid description for
the dielectric function.
Depending on the thickness of the film, the transmission is mainly determined 
by $\epsilon ^{\prime}$~$(\sim 1/\lambda^2)$ or 
$\epsilon ^{\prime \prime}$~$(\sim \sigma_1)$.
For NbN we find a behavior consistent with the BCS-formalism.
For \dbc, results on films of thicknesses ranging 
from 20 to 80 nm are presented. 
We find a T$^2$ dependence of the penetration depth at 
low temperatures and a strongly enhanced conductivity below T$_c$, indicative
of a strong reduction of the quasiparticle scattering.
For the 20 nm film the transmission is 
dominated by $\epsilon ^{\prime \prime }$, even down to 60 K.
\end{abstract}
\vspace{1\baselineskip}
{\it PACS}: 74.25.Gz; 74.25.Nf; 74.76.Bz\\
{\it keywords}: thin films, penetration depth, mm-wave spectroscopy\\
\vskip2pc]
\narrowtext

\section{Introduction}
In the process of establishing the symmetry of the order parameter
of the high temperature superconductors, electrodynamical properties 
such as the penetration depth ($\lambda$) and the
conductivity ($\sigma_1$) are playing a major role.
Most contemporary results on the temperature dependence of $\lambda$
have been obtained at microwave frequencies using a resonant cavity in 
a perturbative mode\cite{Hardy93,Wu,Zhang}.
At temperatures lower than 40 K, Hardy {\em et al.}\cite{Hardy93} 
observed a linear temperature
dependence in a \y695 ~single crystal using a loop-gap resonator at 1 GHz.
This dependence was predicted to exist for a 
superconductor having lines of nodes in the gap function, as for instance 
in case of a d-wave symmetry\cite{Annet}.
The {\em un}conventional pairing symmetry has been confirmed by 
other techniques and properties
\cite{Wollman,Tsuei,Shen,Mason,Deveraux} and it is now 
generally accepted, although consensus has not been reached about
the {\em exact} symmetry.\\
\indent
The linear behavior, believed to be intrinsic, was however initially
never observed in thin films.
Several groups found other 
dependencies ranging from T$^2$ to exponential\cite{Klein,Fiory,Beasley,Ma},
presumably originating from extrinsic sources such as resonant 
impurity scattering or weak links\cite{Annet,Hirschfeld,Halbritter}.
Indeed, in case of a d-wave superconductor, it has been shown theoretically 
that extrinsic scattering sources can turn the linear temperature 
dependence into a T$^2$ dependence\cite{Annet,Hirschfeld}.
De Vaulchier {\em et al.}\cite{Vaulchier} were the first to report
the linear dependence 
in a \yde ~thin film using mm-wave transmission. 
They also noticed a correlation between the observation  
of a large $\lambda_L$ and the T$^2$ dependence, 
using the {\em absolute} information obtained by this technique.
The correlation was interpreted as evidence for 
the extrinsic nature of the nonlinear dependence, for example due to
weak links\cite{Halbritter}.\\
\indent
Other unconventional behavior at mm-wave frequencies has been observed 
in the conductivity $\sigma_1$\cite{Bonnprl,Nuss,Gao}. Unlike the 
reduced $\sigma_1$ expected for a BCS-superconductor, the 
conductivity is largely {\em enhanced} upon entering the superconducting 
state, showing a broad peak at temperatures ranging from 30 to 70 K. 
The peak shows some similarities with a BCS-coherence 
peak\cite{Holczer,KleinGrun}, however, the absence of a Hebel-Slichter 
peak in NMR-data and the strong frequency dependence
of both its amplitude and position are indications that an alternative 
explanation should be used. 
In particular, the maximum in $\sigma_1$ for the cuprates has been
interpreted as originating from 
an anomalously strong reduced scattering rate ($\gamma$) below T$_c$, 
such that the conductivity increases despite  
the decreasing quasiparticle density.
Using Zn-doping in \y695 ~single crystals, Bonn {\em et al.}\cite{Bonnprb} were 
able to show that the maximum in the conductivity is reduced if the disorder
in the material is increased, since the drop in $\gamma$ is then limited by the 
residual scattering.
The strong reduction of $\gamma$ indicates that ordinary 
electron phonon scattering cannot be the {\em main} 
quasiparticle scattering mechanism. 
This is supported further by the linear temperature dependence
of the dc-resistivity in the normal state for the optimally doped
material, indicating that  
the "normal" state does not follow ordinary Fermi liquid predictions.\\
\indent
Measuring thin films using the aforementioned resonant cavity technique
is usually rather complicated, since the only option one has is to
utilize the film as the cavity endplate. This inevitably leads to 
leakage, thereby complicating the analysis. We will show in the course 
of this paper that measuring the mm-wave transmission through a thin film 
can be a good alternative for measuring the electromagnetic properties.
In fact, one is able to obtain {\em absolute} information on both the 
real and imaginary part of the dielectric function\cite{Nagashima}. 
Moreover, by selecting the proper film thickness
we can control the sensitivity such that 
either the conductivity ($\sigma_1$) or the 
superfluid density ($\sim 1/\lambda^2$) dominates the temperature dependence
of the transmission.
We will show that for certain values of the London penetration 
depth $\lambda_L$ and the conductivity $\sigma_1$, 
one can choose the film thickness such that even in the superconducting state
down to low temperatures 
\mbox{($\sim$ 60 K)} the influence of the 
superfluid density is insignificant.

\section{Experimental}
\begin{figure}[htb]
\begin{center}
\leavevmode
\epsfxsize=7.5cm
\epsffile{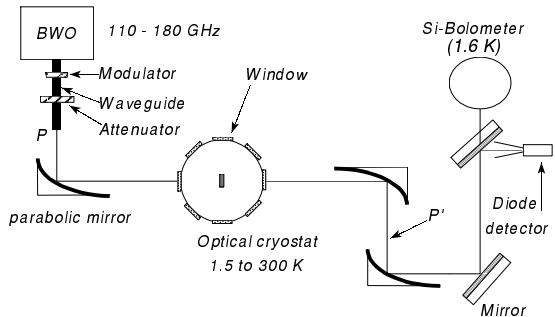}
\end{center}
\caption{{\it Experimental setup}}
\label{expsetup}
\end{figure}

\begin{figure}[htb]
\begin{center}
\leavevmode
\epsfxsize=7.5cm
\epsffile{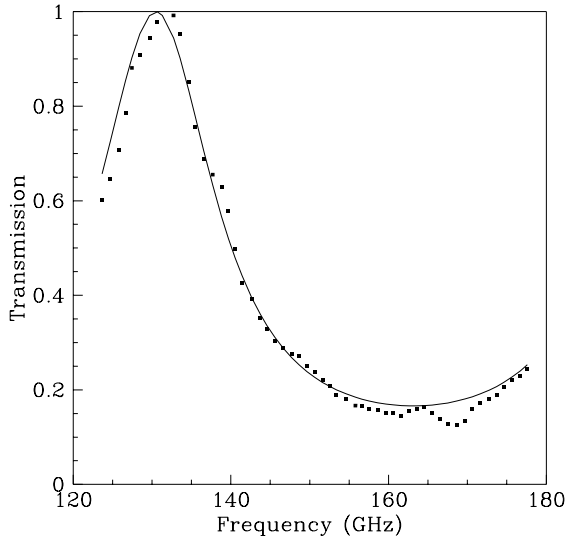}
\end{center}
\caption{The transmission through a bare LaAlO$_3$
substrate at room temperature, showing its
Fabry-Perot resonance. The optical constants
used for the fit were: $n~=~4.70$ and $k~=~0$.}
\label{lao}
\end{figure}

Our experimental setup is shown in fig. \ref{expsetup}. 
As a source we use 
a Backward Wave Oscillator (BWO), having a broad frequency response ranging
from 110 to 180 GHz, making scans as
a function of frequency feasible. The radiation is first coupled into 
a waveguide, through a modulator and a calibrated attenuator. The former
creates 
the ac-response necessary for the detector, while the latter is used to
maintain the signal level comparable during three sequential measurements,
{\em i.e.} thin film on substrate, bare substrate and reference aperture. 
This ensures  
that the response of the detector is linear over the full frequency range.
The radiation is then coupled out of the waveguide
using a Gaussian horn, and treated quasi-optically thereafter. 
We use a parabolic mirror, placed slightly out 
of  focus to make a 3:1 image at the center of the cryostat
(1.5 - 300 K) where the superconducting thin film is situated.
Another parabolic mirror is used to make a second focus (1:1) which
can be utilized for room temperature measurements.
Eventually the beam is focused onto either one of two detectors,
a highly sensitive liquid $^4$He cooled Si-bolometer, 
or a fast waveguide diode detector.\\
\indent
The NbN-film of 55 nm thickness was deposited on an MgO substrate
using DC reactive magnetron sputtering. This was done in a gas 
composition of 3.0\% CH$_4$ / 30.0 \% N$_2$ / Ar, at a temperature of 
approximately 840 $^\circ$C\cite{Barber}. 
T$_c$ was enhanced to 16.5 K, by the inclusion of carbon. 
The resistivity ratio (16.5 K/ 300 K) was close to unity.\\
\indent
The \dbc ~films of thicknesses ranging from 20 to 80 nm, 
were deposited by RF sputtering on LaAlO$_3$ substrates 
using the (100) surface. 
The substrate temperature was 745 $^\circ$C while a mixture
of argon and oxygen gas was used, at pressures of 105 and 45 mTorr 
respectively. After the deposition the samples were annealed in 200 mTorr
oxygen for 30 minutes at 450 $^\circ$C. The quality of the surface was checked
with X-ray diffraction, which showed very good crystallization.
Due to the high crystallinity of the films, the oxygen diffusion process 
was rather slow resulting in a somewhat reduced T$_c$ (88 K) for one of the 
films (20 nm).\\
\indent
During measurements three sequential scans as a function of frequency
are taken at a fixed temperature: thin film on substrate, bare 
substrate and reference aperture. The latter is used to yield {\em absolute}
transmission coefficients for both sample and substrate, which are 
then used in the remainder of the analysis.
The transmission through a two layer system is given by
the Fresnel equations:
\begin{equation}
t~=~\frac{\tau_{02} e^{i\psi} t_{20}}{1~-~r_{20}\rho_{20}e^{2i\psi}}
\hspace{1mm},\hspace{1cm}\psi~=~\frac{2\pi\nu d_{s}p}{c}
\label{fresnel}
\end{equation}
where $\nu$ is the frequency of the incident radiation, p the 
complex index of refraction of the substrate, d$_s$ is the thickness
of the substrate and:
\begin{equation}
\tau_{02}~=~t_{01} e^{i \phi}\frac{1}{1~-~r_{10}e^{i \phi}r_{12}
e^{i \phi}} t_{12}\hspace{1mm}
\label{tau02}
\end{equation}
\begin{equation}
\rho_{20}~=~r_{21}~+~t_{21} e^{2 i \phi} r_{10} \frac{1}{1~-~r_{10}
e^{i \phi}r_{12}e^{i \phi}} t_{12}
\label{rho20}
\end{equation}
Here $\phi=2\pi\nu d_{f}\sqrt{\epsilon}/c$ is a complex phase,
$\epsilon$ is the dielectric function of the film 
and d$_f$ is the film-thickness. The reflection 
and transmission coefficients at each interface are:
\begin{equation}
r_{ij}~=~\frac{n_i-n_j}{n_i+n_j}\hspace{.5cm},\hspace{.5cm} t_{ij}~
=~\frac{2n_i}{n_i+n_j}
\label{intface}
\end{equation}
The multiple reflections in the substrate are incorporated
in the phase factor $e^{i\psi}$, while a similar contribution from the
film is included in a phase factor $e^{i\phi}$, incorporated in the 
transmission and reflection coefficients $\tau_{02}$ and $\rho_{20}$.\\
\indent
First, the interference pattern of the substrate is used
to obtain its full dielectric function ($\tilde{n} = n+ik$) 
at each temperature. In fig. \ref{lao} the transmission of 
a LaAlO$_3$ substrate is shown together with the obtained fit.
Then we continue by modeling the transmission through the sample
using the full Fresnel equations, the measured optical constants of the
substrate and a two-fluid model for 
the dielectric function of the film:
\begin{equation}
\epsilon = \epsilon_{\infty} - \frac{\nu_{pn}^2}{\nu(\nu + 
i \gamma(T))} - \frac{c^2}{(2\pi)^2 \nu^2 \lambda^2(T)}
\label{twofl}
\end{equation}
Here $\nu_{pn}$ is the normal state plasma frequency, $\gamma(T)$ 
is the scattering rate, $\lambda(T)$ is the penetration depth
and $\nu$ is the measurement frequency.
We recall that $\epsilon^{\prime \prime} = 2 \sigma_1/\nu$ above T$_c$, 
and $\sigma_1 \rightarrow 1/\rho_{dc}$ for low frequencies.
Furthermore, $\epsilon^{\prime} = - c^2/(\nu\lambda)^2$ at low temperatures
and is therefore directly related to the superfluid density 
of the material. For the measurements presented in this paper,
the scattering rate $\gamma$ remains larger than the measurement 
frequency $\nu$, so that $\lambda$ and $\sigma_1$ have only a weak
frequency dependency. Mm-wave experiments under conditions 
where $\nu > \gamma$ have been reported by 
D\"{a}hne {\em et al.}\cite{Dahne}.\\ 
\indent
In order to see the influence of  
$\epsilon^{\prime}$ and $\epsilon^{\prime \prime}$ on the transmission
and its temperature dependence, we consider the idealized
but nevertheless instructive case of a free standing film in the limit 
that $|2\pi\nu d\epsilon/c| \ll 1$.
The transmission is then given by
\begin{equation}
T = \left( 1 + (\frac{2 \pi \nu d \epsilon^{\prime \prime}}{c}) 
+ (\frac{\pi \nu d \epsilon^{\prime}}{2c})^2 \right)^{-1}
\label{freefilm}
\end{equation}
where $d$ is the film thickness. 
From this expression it is evident that the 
thickness plays an important role 
in determining the relative significance of the real and imaginary parts. 
For very thin films the 
absorptive term ($\sim \epsilon^{\prime \prime}$) will dominate, 
while for thicker films, the reactive term 
($\sim \epsilon^{\prime}$) will be most significant. 
From eq.\ (\ref{freefilm}) one can estimate a {\em crossover} 
thickness ($d_c$):
\begin{equation}
d_c = \frac{2 c \epsilon^{\prime \prime}}
{\pi \nu (\epsilon^{\prime})^2}
\label{crithi}
\end{equation} 
Since the magnitudes of both contributions $\epsilon^{\prime}$ and
$\epsilon^{\prime \prime}$ are very different for a normal metal or 
a superconductor (due to the presence of the superfluid), 
$d_c$ will be different for the two cases. 
Assuming that $\nu^2 \ll \gamma^2$,
we find that for a metal 
$d_c\approx 2 c\gamma^3 \pi^{-1}(\nu\nu_{pn})^{-2}$, while 
for a superconductor 
$d_c\approx 32\pi^3 \nu_{pn}^2 \nu^2 \lambda^4 c^{-3}\gamma^{-1}$.
Taking common values for 123-superconductors ($\gamma=100~
\mbox{cm}^{-1},~\nu=5~\mbox{cm}^{-1},~\nu_{pn}=
10^4~\mbox{cm}^{-1}~\mbox{and}~\lambda =2000~\AA$), 
we obtain that $d_c \sim 2.5\,\mu$m in the normal 
state, while $d_c \sim 0.4\,\AA$ for T$ \ll \mbox{T}_c$.
Consequently, for most thin films the 
transmission in the normal state will be determined
by $\sigma_1$, while in the superconducting state 
it will be determined by $\lambda$.
Equivalently, by fixing the thickness and using the temperature dependence of 
the sample properties one can estimate the temperature range at which both 
contributions will be equally important.
This will be shown in a more specific analysis 
using the measured values for \dbc, given in the following section.

\section{Results}
\begin{figure}[htb]
\begin{center}
\leavevmode
\epsfxsize=7.5cm
\epsffile{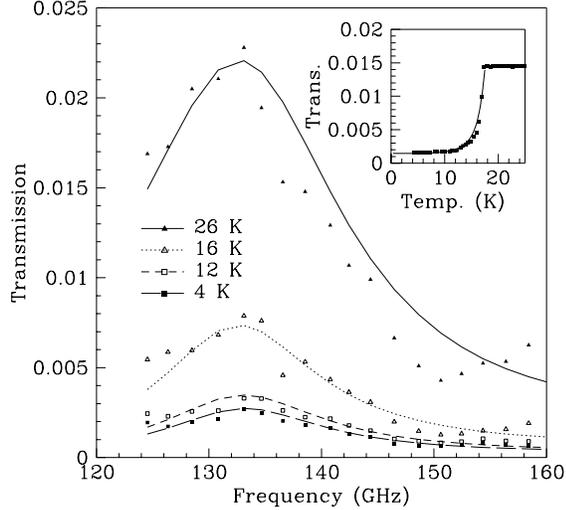}
\caption{Transmission through NbN on MgO (thickness = 55 nm, T$_c$ = 16.5 K) 
for 4 different temperatures together with their fit 
(4 K: solid squares, 12 K: open squares, 16 K: open 
triangles and 26 K: solid triangles). Inset:
temperature dependence at 140 GHz together with a BCS-fit using 
$\sigma_1$(17 K)~=~15000 $\Omega^{-1}$cm$^{-1}$ and $\lambda_L$~=~400 nm.}
\label{nbn}
\end{center}
\end{figure}

\begin{figure}[htb]
\begin{center}
\leavevmode
\epsfxsize=7.5cm
\epsffile{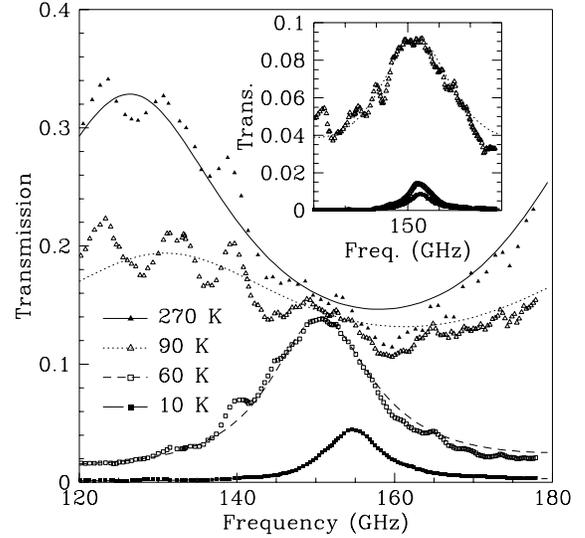}
\caption{Transmission through \dbc ~on LaAlO$_3$ 
(thickness = 20 nm, T$_c$ = 88 K) for 4 different temperatures 
together with their fit (10 K: solid squares, 60 K: open squares,
90 K: open triangles and 270 K: solid triangles).
Inset: Transmission of \dbc ~(d = 34 nm) for three different temperatures
(10, 60 and 90 K).}
\label{dbcotra}
\end{center}
\end{figure}

\begin{figure}[htb]
\begin{center}
\leavevmode
\epsfxsize=7.5cm
\epsffile{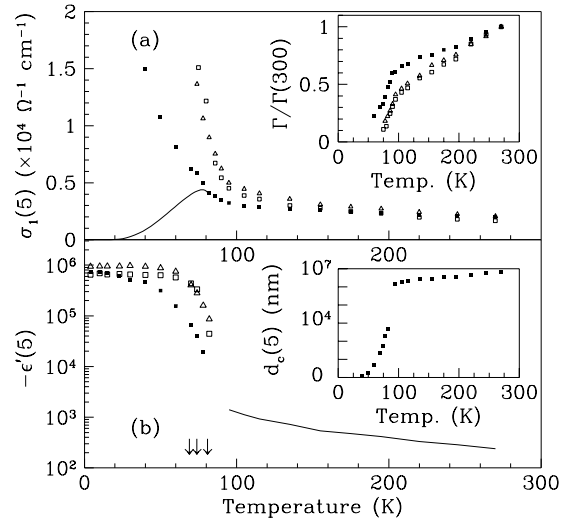}
\caption{a: Temperature dependence of $\sigma_1$  
for three \dbc ~films with different
thicknesses (20 nm: solid squares, 34 nm: open triangles and 80 nm:
open squares). The normalized conductivity for a BCS-superconductor
having a T$_c$ of 92 K is included as the solid line.
The inset shows $\gamma$, normalized to its RT-value.
b: Temperature dependence of $\epsilon^{\prime}$. 
The arrows indicate the temperatures where 
the critical thickness, $d_c$ (inset), is approximately equal to the 
film thickness (20, 34 and 80 nm from left to right). All
quantities have been calculated at the center frequency, $\nu~=~5~cm^{-1}$.}
\label{sigeps}
\end{center}
\end{figure}

\begin{figure}[htb]
\begin{center}
\leavevmode
\epsfxsize=7.5cm
\epsffile{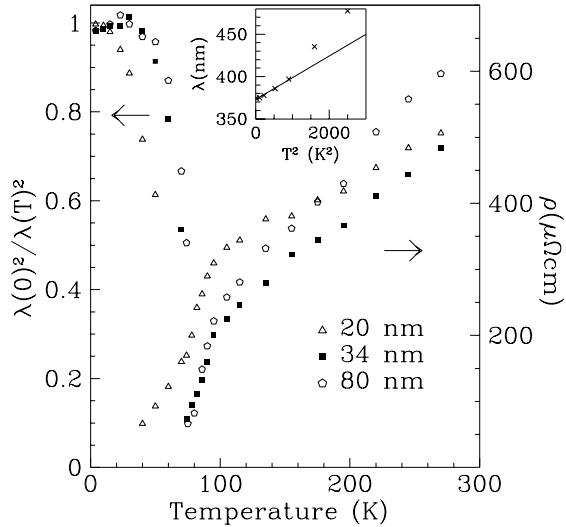}
\caption{Temperature dependence of the resistivity (right hand side) and 
the superfluid density $\lambda(0)^2/\lambda(T)^2$
(left hand side) for 
three \dbc ~films with different thicknesses. 
(20 nm: triangles, 34 nm: squares and 80 nm:
diamonds). In the inset the temperature dependence of the 
penetration depth is shown for the 20 nm film. The open symbols are obtained
using approach A, while the close symbols result from approach B.}
\label{rhons}
\end{center}
\end{figure}

In order to check the reliability of the technique we began by
first measuring 
a conventional superconductor (NbN) to make sure that we would be able 
to resolve the intrinsic properties of the thin film.
In fig. \ref{nbn} the transmission of the NbN film on an MgO substrate 
is shown as a function
of frequency, for several temperatures, both above and below T$_c$.
The peak in the interference spectrum is determined by the MgO substrate,
having $n = 3.43$ and $k = 0$. We are able to fit the spectra 
using the Drude description of equation (\ref{twofl}).
As expected for a metallic film, $\epsilon^{\prime \prime} \gg
|\epsilon^{\prime}|$, while in the superconducting state the opposite 
is valid, $|\epsilon^{\prime}| \gg \epsilon^{\prime\prime}$. 
For the fit we have focused our attention on the main peak around 135 GHz.
Notice, furthermore, that the interference pattern shows no major changes 
besides the reduced amplitude in the superconducting state.\\
\indent
Plotting the temperature dependence at one particular frequency (140 GHz) 
as shown in the inset of fig.~\ref{nbn}, the dramatic change in 
transmission at T$_c$ is more easily observed.
The temperature dependence of the transmission coefficient 
can be fitted very well
using the assumptions that $\lambda$(T) can be described by the 
Gorter-Casimir relation\cite{Tinkhamint},
$\sigma_1$(T) follows the Mattis-Bardeen relations\cite{Mattis}
and \mbox{$2 \Delta$~/~k$_B$T~=~4.0.} 
The only adjustable parameters used in the calculation were the normal state 
conductivity ($\sigma_1\,(17 K)\,=\,15000~\Omega^{-1}\mbox{cm}^{-1}$) 
which determines the absolute transmission at 17 K, and 
the London penetration depth ($\lambda_L$~=~400 nm) which determines the 
transmission coefficient at low temperatures.
Although this penetration depth is somewhat higher than the lowest reported
values of about 100 nm\cite{Shoji}, there is considerable 
variation in the literature
due to a wide range of film structures.\\
\indent
In fig.~\ref{dbcotra} the transmission through a 20nm thick \dbc ~film 
on a LaAlO$_3$ substrate
is shown. 
Similar results as presented below have been reproduced 
on a different batch of films.
The interference pattern of the substrate 
at room temperature was shown already in fig. \ref{lao}. 
From this we obtained n = 4.70 and k = 0. As expected,
the optical constants are temperature independent for the perovskite
substrate.
The additional oscillations present in the transmission are caused by 
internal reflections within the sampleholder. Due to the 
slightly modified standing wave pattern, these do not cancel completely 
after division by the transmitted signal through a reference hole.\\
\indent
In contrast to the results for the NbN film, 
the interference pattern of \dbc~thin film changes
dramatically when the 
temperature is lowered. 
This is due to the stronger change in conductivity 
of the film,
thereby altering the matching of the impedances of both film and substrate.
\begin{figure}[htb]
\begin{center}
\leavevmode
\epsfxsize=7.5cm
\epsffile{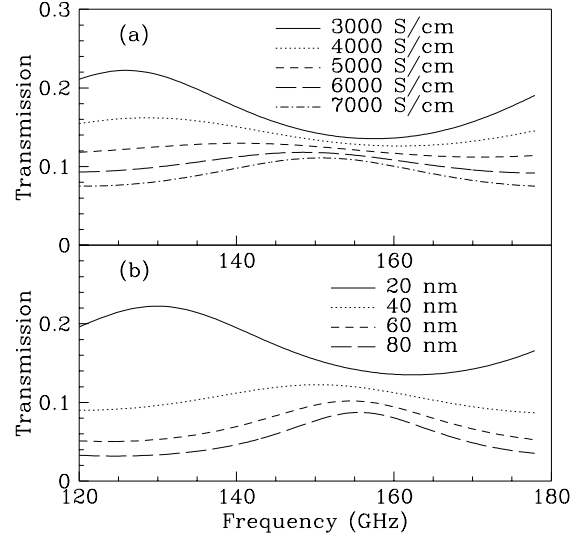}
\caption{Calculated transmission through a layered system (thin film
on a substrate) as a function of conductivity (a), or thickness (b).
For the substrate, the optical constants of LaAlO$_3$ were used
and the conductivity for (b) was 3000 $\Omega^{-1}$cm$^{-1}$.}
\label{simul}
\end{center}
\end{figure}
This effect can be demonstrated by modeling the 
transmission through a similar stratified
system, changing either the optical conductivity $\sigma_1$ 
of the film, keeping the thickness fixed at 20 nm (fig. \ref{simul}a), or
by altering its 
thickness at a fixed conductivity of 3000 $\Omega^{-1}$cm$^{-1}$ 
(fig. \ref{simul}b). 
Both parameters will have 
a similar effect on the transmission since this is mainly 
affected by their product.
For a certain set of parameters the 
interference effect disappears completely, showing
that the light passes through the substrate only once. Additionally, the 
magnitude of the dielectric function matching this requirement
determines the absolute value of transmission at the turning point.
This model calculation illustrates furthermore that one has to be careful
with the interpretation of transmission curves 
measured as a function of temperature
at a fixed frequency. A curve taken at 130 GHz will show a much 
larger temperature dependence than a curve taken at 150 GHz, although
they will yield the same intrinsic film properties, once interference effects 
are taken into account. \\
\indent
At 150 GHz the 
transmission is nearly constant at temperatures higher than 60 K. At
lower temperatures the amplitude of the interference peak 
drops and its width decreases rapidly.
The transmission for the 34 nm film is depicted in the inset of 
fig. \ref{dbcotra} for the same temperatures as the 20 nm film
(90, 60 and 10 K).
We see that the reduction in transmission at intermediate temperatures
is much larger in the 
thicker film. As we will argue below, this is due to the stronger 
contribution of the superfluid in the 34 nm film.
The transmission data for the 80 nm film
show qualitatively similar behavior.\\
\indent
To fit the transmission data presented in \mbox{fig. \ref{dbcotra}} we use
two different approaches. First we start in 
the normal state, knowing that there is no superfluid
fraction present (approach A). This yields both $\epsilon{^\prime}$ 
and $\epsilon^{\prime \prime}$, 
where the sensitivity in the metallic case is most extensive 
for $\epsilon^{\prime \prime}~(\sim \nu_{pn}^2/\gamma)$.
We continue this approach even when the sample is cooled below 
T$_c$ where the addition of a superfluid contribution doesn't 
influence the fit for the 20 nm film significantly.
We proceed until the peak becomes too narrow and 
fitting is no longer possible.
The second approach (B) is to start at the lowest possible 
temperature (5 K) and assume that
the superfluid fraction is the dominant contribution.
We proceed to higher temperatures until the peak starts to broaden and 
the maximum remains nearly constant, thus inhibiting 
fitting with the superfluid as the only contribution.
Obviously there will be a temperature range for 
which both terms are comparable,
in which case a quantitative description is more complicated.\\
\indent
In fig. \ref{sigeps}a $\sigma_1$ 
is shown, while in fig. \ref{sigeps}b the total $\epsilon^{\prime}$, including 
the superfluid contribution, has been depicted. 
Both quantities have been 
determined at the center frequency, $\nu = 5~\mbox{cm}^{-1}$,
using $\sigma_1~=~\nu_{pn}^2\gamma(\nu(\nu^2~+~\gamma^2))^{-1}$ and 
$\epsilon^{\prime}~=~-\nu_{pn}^2(\nu^2~+~\gamma^2)^{-1}-
c^2(2\pi \nu \lambda)^{-2}$. 
To determine $\sigma_1$ we used the values for 
$\nu_{pn}$ and $\gamma$ obtained from
the fit following approach A. The solid curve in fig. \ref{sigeps}b 
corresponds to an estimate of $\epsilon'$ based on the expression
$\epsilon'=-2\sigma_1/\gamma$ which is valid for the Drude model. 
In the inset of the upper panel the normalized scattering rate is 
depicted.
Below T$_c$ $\sigma_1$ exhibits 
a rapid increase for all films. For comparison, 
the conductivity for a BCS-superconductor with a T$_c$ of 92 K
is included in
the upper panel (solid line). The conductivity was normalized to the value 
of $\sigma_1$ for \dbc~at 92 K. 
This emphasizes the strikingly different
behavior of the high T$_c$ superconductor. Similar behavior 
has been observed 
before in \yde\cite{Bonnprl,Nuss}, and is taken to be evidence 
of a rapidly suppressed scattering rate $\gamma$ below T$_c$.
Since, within the model used, $\sigma_1 \sim \nu_{pn}^2/\gamma$, 
two competing effects determine its temperature dependence.
Below T$_c$ the density of normal carriers will be reduced
thereby reducing the plasma frequency, $\nu_{pn}$,
while the scattering of 
quasiparticles will also be reduced when the temperature is lowered.
Having two different temperature dependencies produces a maximum
in the conductivity. This maximum also resembles, superficially,
a BCS coherence peak but shows   
a different temperature and frequency behavior.\\
\indent
The total dielectric response,
$\epsilon^{\prime}$, is mainly determined by the 
superfluid contribution. For the 20 nm film $\epsilon^{\prime}$
can not be determined accurately from 50 to 90 K, caused by the 
insensitivity of the transmission to $\nu_{ps}$ in this range.\\
\indent
Knowing the approximate values for the dielectric properties 
we return to equation (\ref{crithi}) in order to show that 
the earlier claim
that the sensitivity shifts from $\sigma_1(T)$ to $\lambda(T)$ in this
temperature range was valid. 
Using equation (\ref{crithi}) and the measured values for $\sigma_1$ and 
$\epsilon^{\prime}$ we can calculate the critical thickness.
The result can be seen in the inset in the lower panel of fig \ref{sigeps}.
Since the values used to calculate d$_c$ are {\em intrinsic} material 
properties the curve looks similar when we use 
the dielectric properties obtained for the 34 and the 80 nm film.  
We can hence estimate at which temperature the critical thickness 
is approximately equal to the film thickness. These temperatures 
have been indicated in fig \ref{sigeps}b by three arrows, where the 
thinnest film is represented by the lowest temperature.
Around these temperatures both approaches A and B are inefficient,
resulting in a larger uncertainty in the obtained absolute values 
of both $\sigma_1$ and $\epsilon^{\prime}$.\\
\indent
More results of the fitting procedure are displayed in 
fig. \ref{rhons} for all three films. 
The resistivity 
$\rho$ is shown on the right hand side, while on the left hand side of 
fig. \ref{rhons} the superfluid density 
($\lambda(0)^2/\lambda(T)^2$) is plotted.
The results at low temperatures have been obtained using approach B. 
At higher temperatures $\rho$ has been calculated 
by direct inversion of $\sigma_1$ obtained using approach A, assuming
that $\sigma_2$ can be neglected. The dc-resistivity obtained 
on samples prepared under identical conditions showed
$\rho(300)=650\,\mu\Omega cm$ and $\rho(T_c)=230\,\mu\Omega cm$,
which is in good agreement with the mm-wave data in fig \ref{rhons}.
In the normal state we see a linear temperature dependence of the 
resistivity, similar to the dc-behavior.
However, the slope tells us that there is a fairly large residual scattering.
For instance for the 20 nm film 
the intercept at T~=~0 K is about 225 $\mu\Omega$cm. 
Moreover, the slope is
about twice as large as values measured for single crystals 
($1.05~\mu\Omega$cm/K vs. $0.45~\mu\Omega$cm/K), indicating 
that the difference cannot be explained by merely adding a temperature 
{\em in}dependent residual resistivity term.\\
\indent
The absolute penetration depth of the 20 nm film is shown in the inset
of fig. \ref{rhons}. The penetration depth shows a T$^2$-dependence
at lower temperatures and a rather large $\lambda_L$ of 370 nm. The other
films show similar behavior with a slightly different 
zero temperature penetration
depth (325 and 360 nm for the 34 and 80 nm films respectively).
Similar behavior was observed before by de 
Vaulchier {\em et al.}\cite{Vaulchier} and was taken to be evidence
for the extrinsic nature of the temperature dependence, 
due to the existence of weak links within the film.
This provides additional evidence for the d-wave scenario, where 
the power-law temperature dependence of $\lambda$ is lifted to a higher
order by the presence of additional scattering.
The slope of the quadratic curve ($\sim 0.025~\AA$/K$^2$) 
is about the same as that reported for one 
of the films in ref.\cite{Vaulchier}, although due to the extrinsic nature
of the phenomenon there is no need for these values to be the same. 

\section{Conclusions}
In summary we can say that mm-wave transmission can be used as a complimentary
technique to characterize and study superconducting thin films on a 
fundamental level. From the NbN-data we see that we are able to resolve the
intrinsic behavior and obtain absolute values for both the real and 
imaginary part of the dielectric function and deduce values for 
$\sigma_1~(15000~\Omega^{-1}\mbox{cm}^{-1}$) and $\lambda_L~(400~\mbox{nm})$. 
The temperature dependence follows the expected BCS-behavior.\\
\indent
Furthermore we have studied the transmission through 
\dbc ~films of different thickness (20, 34 and 80 nm). 
We observed an enhanced conductivity in going into the superconducting
state, indicative of a large reduction in the scattering rate $\gamma$
just below T$_c$.
For the resistivity in the normal state, we find the linear behavior
typical for the cuprates. From the intercept 
at T~=~0~K we learn that there is an additional residual
scattering in the film, probably due to the same oxygen deficiency
that reduces T$_c$ slightly. The London penetration depth is fairly 
large for all films (325 - 370 nm), 
and has a T$^2$ dependence, consistent with a d-wave symmetry picture
plus an additional extrinsic scattering source.
For the thinnest film (20 nm), the superfluid density has no influence  
on the transmission coefficient down to temperatures well below $T_c$. 
Therefore the temperature dependence of the transmission is completely 
determined by $\sigma_1$, even at temperatures down to 60 K.
The thicker films show a more conventional behavior, where 
$\epsilon^{\prime}$ indeed dominates the transmission in most of the 
superconducting range.\\
\indent
{\it Acknowledgements} We gratefully acknowledge stimulating 
discussions with W. N. Hardy and 
the technical support of C. J. Bos and W. A. Schoonveld.

\end{document}